\documentclass[preprint,sort&compress,11pt,5p]{elsarticle}

\usepackage{moreverb}
\usepackage{amsmath}
\usepackage{algorithmic}
\usepackage{graphics}
\usepackage{graphicx}
\usepackage{subfigure}
\usepackage{enumitem}
\usepackage{color}
\usepackage{framed}
\usepackage{wrapfig}
\usepackage{bm}
\usepackage{mathrsfs}
\usepackage{amsfonts}
\usepackage{multirow}
\usepackage{longtable}
\usepackage{subfigure}
\usepackage{hyperref}
\usepackage{paralist}
\usepackage{indentfirst}
\usepackage{relsize}
\usepackage{extarrows}
\usepackage{lineno,xcolor}
\usepackage{upgreek}
\usepackage{diagbox}
\usepackage{bm}
\usepackage{xcolor}
\usepackage{booktabs}
\usepackage[flushleft]{threeparttable}
\usepackage[linesnumbered,lined,ruled]{algorithm2e}

\graphicspath{ {./} }

\newcommand{\eref}[1]{(\ref{#1})}

\hypersetup{
bookmarks=true,
bookmarksopen=true,
bookmarksnumbered=true,
unicode=false,
pdftoolbar=true,
pdfmenubar=true,
pdffitwindow=false,
pdfstartview={FitH},
pdftitle={My title},
pdfauthor={Author},
pdfsubject={Subject},
pdfcreator={Creator},
pdfproducer={Producer},
pdfkeywords={keywords},
pdfnewwindow=true,
colorlinks=true,
linkcolor=blue,
citecolor=blue,
filecolor=magenta,
urlcolor=blue
}

\journal{Theoretical and Applied Mechanics Letters}

\begin{document}
	
\title{\LARGE Physics-informed deep learning for incompressible laminar flows}

\author[NU1]{Chengping~Rao}
\author[NU2,MIT]{Hao~Sun}
\ead{h.sun@northeastern.edu}
\author[NU1]{Yang~Liu\corref{cor}}
\ead{yang1.liu@northeastern.edu}

\cortext[cor]{Corresponding author. Tel: +1 617-373-8560}

\address[NU1]{Department of Mechanical and Industrial Engineering, Northeastern University, Boston, MA 02115, USA}
\address[NU2]{Department of Civil and Environmental Engineering, Northeastern University, Boston, MA 02115, USA}
\address[MIT]{Department of Civil and Environmental Engineering, MIT, Cambridge, MA 02139, USA}

\begin{abstract}
	\small
Physics-informed deep learning has drawn tremendous interest in recent years to solve computational physics problems, whose basic concept is to embed physical laws to constrain/inform neural networks, with the need of less data for training a reliable model. This can be achieved by incorporating the residual of physics equations into the loss function. Through minimizing the loss function, the network could approximate the solution. In this paper, we propose a mixed-variable scheme of physics-informed neural network (PINN) for fluid dynamics and apply it to simulate steady and transient laminar flows at low Reynolds numbers. {\color{black} A parametric study indicates that the mixed-variable scheme can improve the PINN trainability and the solution accuracy.} The predicted velocity and pressure fields by the proposed PINN approach are also compared with the reference numerical solutions. Simulation results demonstrate great potential of the proposed PINN for fluid flow simulation with a high accuracy.
\end{abstract}

\begin{keyword}
	\small
	Physics-informed neural networks (PINN) \sep deep learning \sep fluid dynamics \sep incompressible laminar flow
\end{keyword}

\maketitle

\section{Introduction}\label{sintro}

Deep learning (DL) has attracted tremendous attentions in recent years in the field of computational mechanics due to its powerful capability in nonlinear modeling of complex spatiotemporal systems. According to a technical report \cite{lee2018basic} by U.S. Department of Energy, a DL-based approach should be featured with the domain-aware, interpretable and robust to be a general approach for solving the science and engineering problems. Recent studies of leveraging DL to model physical system include, just to name a few, \cite{ZHANG201955, yang2016data, tompson:2017, raissi2019physics, raissi2019deep}. These applications can be categorized into two types based on how the DL model is constructed: in either data-driven or physics-informed manner. In a data-driven framework, the DL model is constructed as a black-box to learn a surrogate mapping from the formatted input $\mathbf{x}\in\mathbb{R}^{m}$ to the output $\mathbf{y}\in\mathbb{R}^{n}$. The exceptional approximation ability of deep neural network (DNN) makes possible to learn the mapping even when the dimensionality $m$ and $n$ are very high. The training dataset $\{\mathbf{x, y}\}$, typically very rich, can be obtained by conducting high-fidelity simulations using exact solvers (e.g., see \cite{yang2016data, tompson:2017,geneva2019quantifying}). Nevertheless, obtaining a rich and sufficient dataset from simulations for training a reliable DL model is computationally expensive and requires careful case design. To address this fundamental challenge, physics-informed DL explicitly embed the physical laws (e.g., the governing partial differential equations (PDEs), initial/boundary conditions, etc.) into the DNN, {\color{black}constraining the network's trainable parameters within a feasible solution space}. The objective of exploiting physical laws in DNN is assumed to (1) reduce the large dependency of the model on available dataset in terms of both quality and quantity, and (2) improve the robustness and interpretability of the DL model. In this regard, DNN essentially has the capacity of approximating the latent solutions for PDEs \cite{lagaris:1998, lagaris:2000}, with distinct benefits summarized as follows: (1) the superior interpolation ability of DNN, (2) the approximated solution has a close form with its infinite derivative continuous, and (3) state-of-the-art hardware advances make the numerical implementation and parallelization extremely convenient.

 
\begin{figure*}[t!]
	\centering
	\includegraphics[width=0.9\textwidth]{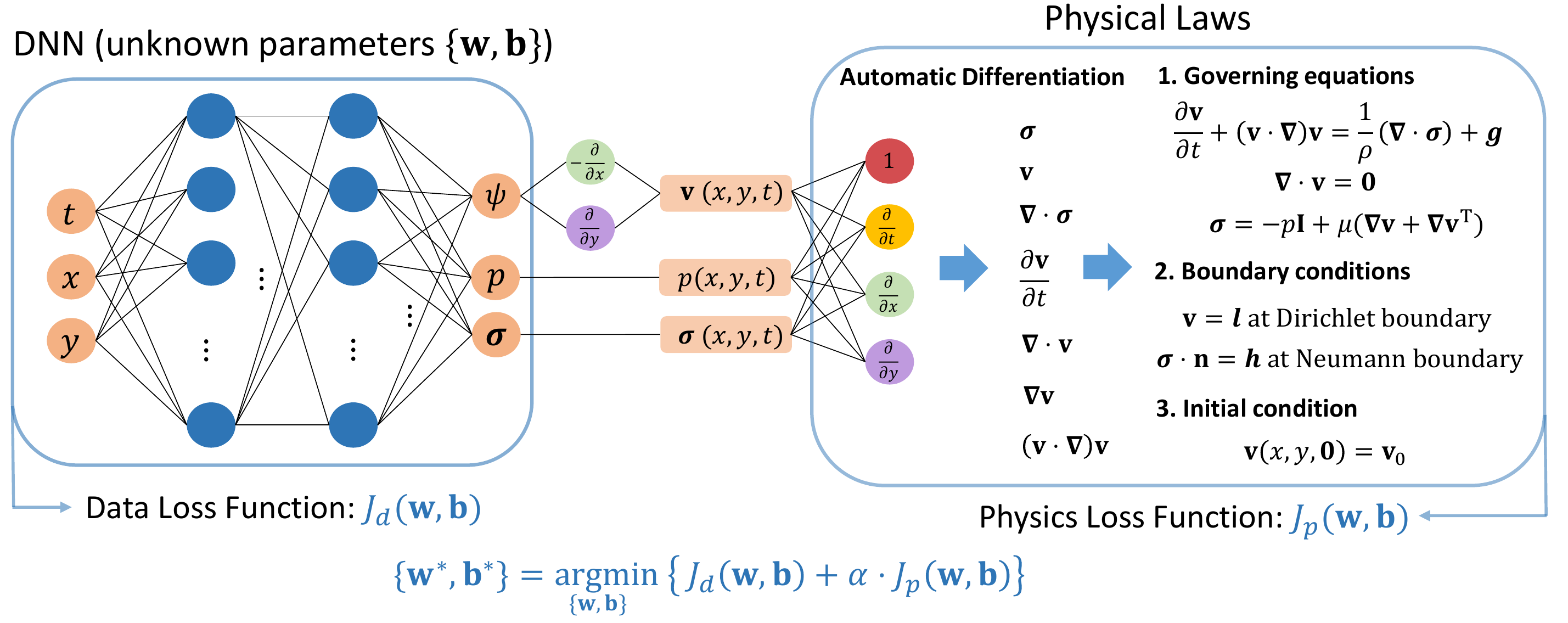}
	\vspace{-6pt}
	\caption{Architecture of the physics-informed neural network for fluid dynamics. Note that $\alpha$ is a user-defined weighting coefficient. $\mathbf{w}$ and $\mathbf{b}$ are weights and biases for the DNN. The constraint of initial and boundary conditions can be converted as residuals adding to the loss function based on Lagrangian multipliers. The data loss $J_d$ is present only when data is available.}
	\label{scheme}
\end{figure*}

More recently, Raissi \emph{et al.} \cite{raissi2019physics, raissi2019deep} introduced a general framework of PINN and demonstrated its capacity in modeling complex physical systems such as solving/identifying PDEs. A huge difference from some of the previous studies is that, in addition to the physical laws, the PINN can also exploit the available measurement data, making it possible to discover the systems whose physics are not fully understood. In particular, seminal contributions of using PINN to model fluid flows have been made recently. For example, Kissas \emph{et al.} \cite{kissas2020machine} employed PINN to predict the arterial blood pressure based on the MRI data and the conservation laws. Sun \emph{et al.} \cite{sun2019surrogate} proposed a PINN approach for surrogate modeling of fluid flows without any simulation data. Zhu \emph{et al.} \cite{zhu2019physics} proposed a physics-constrained convolutional encoder-decoder network and a generative model for modeling of stochastic fluid flows. 

In this paper, we formulate a mixed-variable PINN scheme for simulation of viscous incompressible laminar flows without any measurement data. The remaining this paper is organized as follows. In Section \ref{method}, we will introduce the methodology of the mixed-variable PINN and the mathematical formulation for fluid dynamics. Subsequently in Section \ref{results}, the steady and transient laminar flows passing a circular cylinder will be modeled using the proposed PINN scheme without any simulation data. {\color{black}A comparison study is made to demonstrate the improved solution accuracy and network trainability by the proposed scheme.} Section \ref{conc} summarizes the conclusions. 

\section{Methodology}\label{method}

Let us consider the incompressible Newtonian flow governed by the following Navier–Stokes equations:
\begin{equation} 
\label{maximization} 
\begin{aligned}
\mathbf{\nabla}\cdot\mathbf{v}&=0
\end{aligned}
\end{equation}
\begin{equation} 
\label{nseqn} 
\begin{aligned}
\frac{\partial \mathbf{v}}{\partial t}+(\mathbf{v}\cdot\mathbf{\nabla})\mathbf{v}&=-\frac{1}{\rho}\mathbf{\nabla}p+\frac{\mu}{\rho}\nabla^2\mathbf{v}+\mathbf{g}
\end{aligned}
\end{equation}
where $\nabla$ is the Nabla operator, $\mathbf{v}=(u,~v)$ is the velocity vector, $p$ is the pressure, $\mu$ is the viscosity of the fluid, $\rho$ is the density of fluid and $\mathbf{g}$ is the gravitational acceleration. When leveraging PINN to solve the aforementioned PDEs, minimizing a complex residual loss resulted from Eq. \eref{nseqn} is intractable due to its complex form with multiple latent variables (e.g., $\mathbf{v}$ and $p$) and high-order derivatives (e.g., $\nabla^2$). In order to design an easily trainable PINN, we convert the Navier–Stokes equation in Eq. \eref{nseqn} to the following continuum and constitutive formulations:
\begin{equation} 
\label{continuumeqn} 
\begin{aligned}
\frac{\partial \mathbf{v}}{\partial t}+(\mathbf{v}\cdot\mathbf{\nabla})\mathbf{v}&=\frac{1}{\rho}\mathbf{\nabla}\cdot\bm{\sigma}+\mathbf{g}
\end{aligned}
\end{equation}
\begin{equation} 
\label{consteqn} 
\begin{aligned}
\bm{\sigma}&=-p\mathbf{I}+\mu(\nabla\mathbf{v}+\nabla\mathbf{v}^\text{T})
\end{aligned}
\end{equation}
where $\bm{\sigma}$ is the Cauchy stress tensor and {\color{black} $p=-\mathrm{tr}(\bm{\sigma})/2$}. The benefits of using the continuum-mechanics-based formulation are two-fold: (1) reducing the order of derivatives when a mixed-variable scheme in PINN is used, and (2) improved trainability of DNN as found in the comparison of numerical results. 

The proposed mixed-variable scheme is used in this paper to solve the aforementioned PDEs (see Eqs. \eref{maximization}, \eref{continuumeqn} and \eref{consteqn}) that govern the laminar flow dynamics. The salient feature of PINN is that the physical fields are approximated globally by a DNN. In particular, the DNN maps the spatiotemporal variables $\{t, \mathbf{x}\}^T$ to the mix-variable solution $\{\psi,~p,~\bm{\sigma}\}$, where the stream function $\psi$ is employed rather than the velocity $\mathbf{v}$ to ensure the divergence free condition of the flow. In this way, the continuity equation will be satisfied automatically. For a two-dimensional problem, the velocity components can be computed by $\left [u,v,0 \right ]=\mathbf{\nabla}\times[0,0,\psi]$. Note that $\mathbf{v}=\left [u,v\right]$ is taken as the latent variable. The automatic differentiation is used to obtain the partial derivatives of the DNN output regarding the time and space (e.g., $t$, $x$ and $y$). {\color{black}The loss function is composed of the data loss $J_d$ (if measurements are available) and the physics loss $J_p$. The physics loss $J_p$ is the summation of the governing equation loss $J_g$ and the initial and boundary condition loss $J_{i/bc}$, given by
\begin{equation} 
\label{lossfunpde1} 
J_g = \frac{1}{N_g}\sum_{i=1}^{N_g} {||\bm{r}_g(\mathbf{x}^i, t^i)||^2}
\end{equation}
\begin{equation} 
\label{lossfunpde2} 
\begin{aligned}
J_{i/bc} &= \frac{1}{N_{I}}\sum_{i=1}^{N_{I}} {||\bm{r}_{I}(\mathbf{x}^i, 0)||^2} \\ &+ \frac{1}{N_{nb}}\sum_{i=1}^{N_{nb}} {||\bm{r}_{nb}(\mathbf{x}^i, t^i)||^2} \\&+ \frac{1}{N_{db}}\sum_{i=1}^{N_{db}} {||\bm{r}_{db}(\mathbf{x}^i, t^i)||^2}
\end{aligned}
\end{equation}
where $\bm{r}_{(\cdot)}$ denotes the residual, $||\cdot||$ denotes the $\ell_2$ norm and $N_{(\cdot)}$ denotes the number of collocation points (subscripts  $g$ for governing equation, $I$ for initial condition, $nb$ for Neumann boundary, and $db$ for Dirichlet boundary). The total physics loss $J_p$ is defined as
\begin{equation} 
\label{phyloss} 
\begin{aligned}
J_p = J_g+\beta J_{i/bc}
\end{aligned}
\end{equation}
where $\beta>0$ is a user-defined weighting coefficient for initial and boundary condition loss.} Noteworthy, having the measurement data makes the fluid flow modeling data-driven, which is however not a prerequisite. The architecture of the proposed PINN for fluid dynamics simulation is presented in Fig. \ref{scheme}. In this paper, no measurement data from simulations or physical experiments is used for training the PINN.



\section{Results}\label{results}

\begin{figure}[t!]
	\centering
	\includegraphics[width=0.4\textwidth]{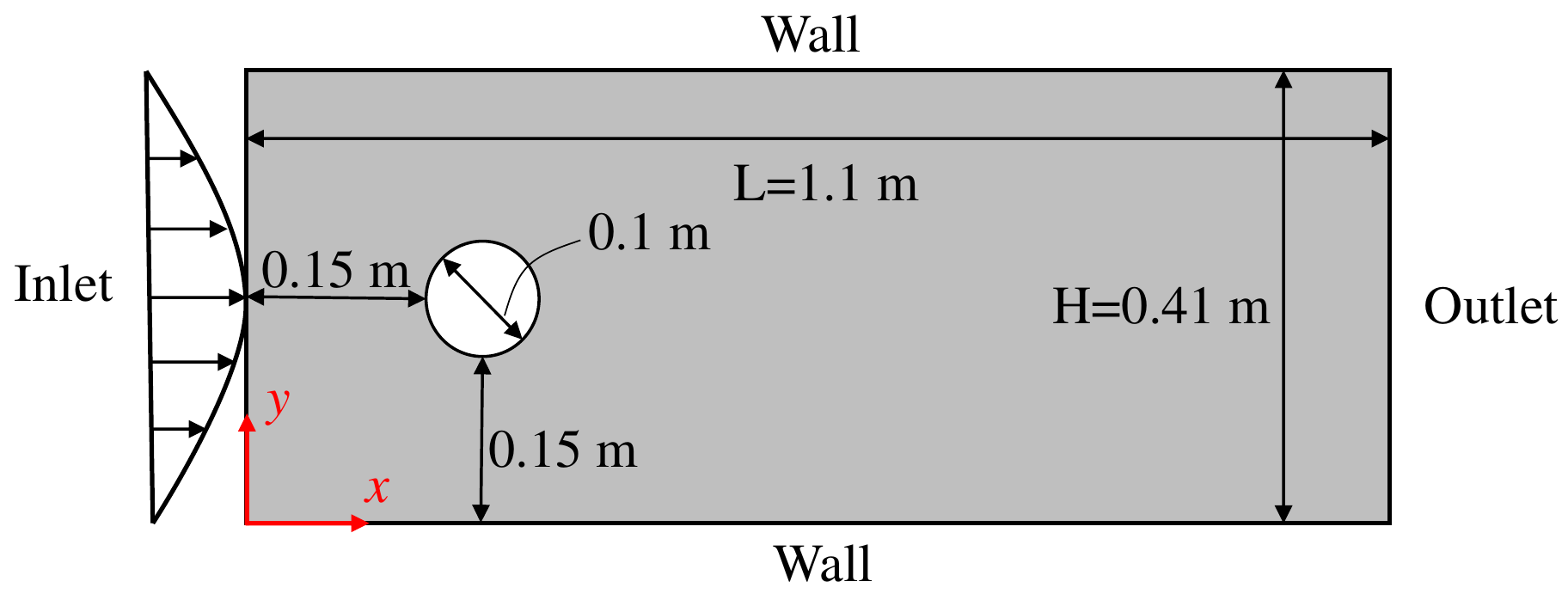}
	\caption{Diagram of the computation model.}
	\label{diagram}
\end{figure}

In this section, we employ the proposed PINN to model the steady and transient flows passing a circular cylinder. A parabolic velocity profile is applied on the inlet while the zero pressure condition is applied on the outlet, as shown in Fig. \ref{diagram}. Non-slip conditions are enforced on the wall and cylinder boundaries. The gravity is ignored in both two cases. The proposed PINN is implemented on the TensorFlow \cite{abadi2016tensorflow} {\color{black} and the source codes can be found in https://github.com/Raocp/PINN-laminar-flow.}

\begin{figure*}[t!]
\centering
\subfigure[]{
\centering
\includegraphics[width=0.25\linewidth]{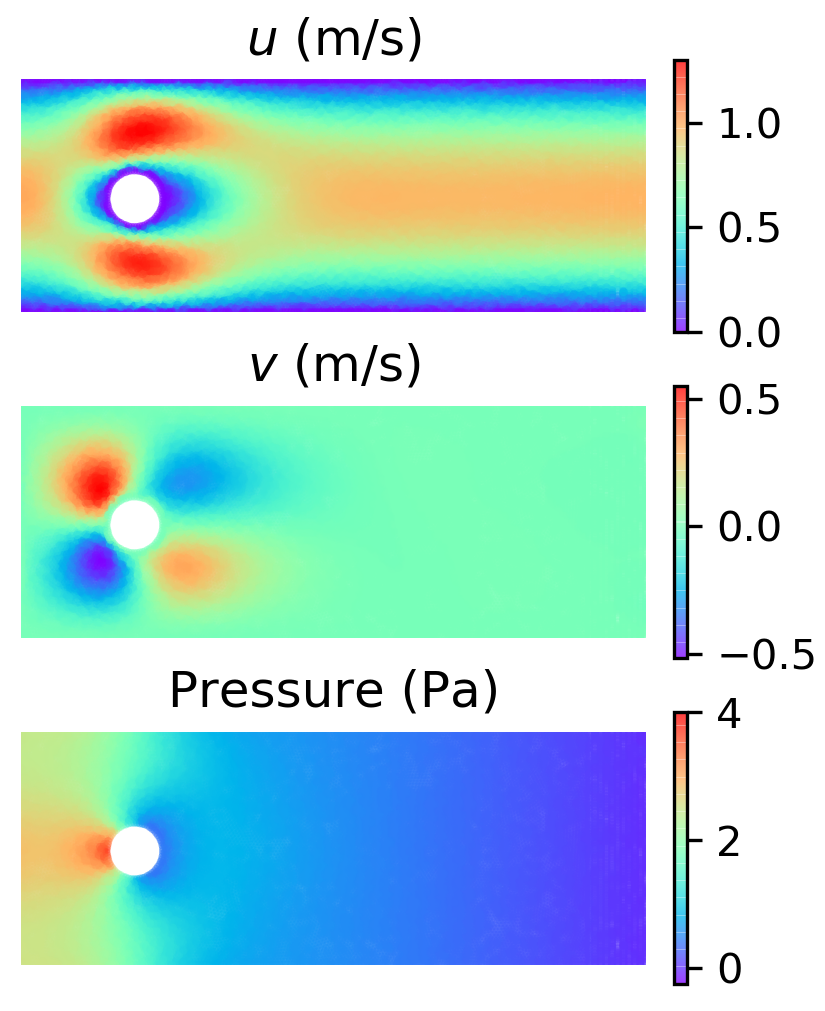}
}%
\subfigure[]{
\centering
\includegraphics[width=0.25\linewidth]{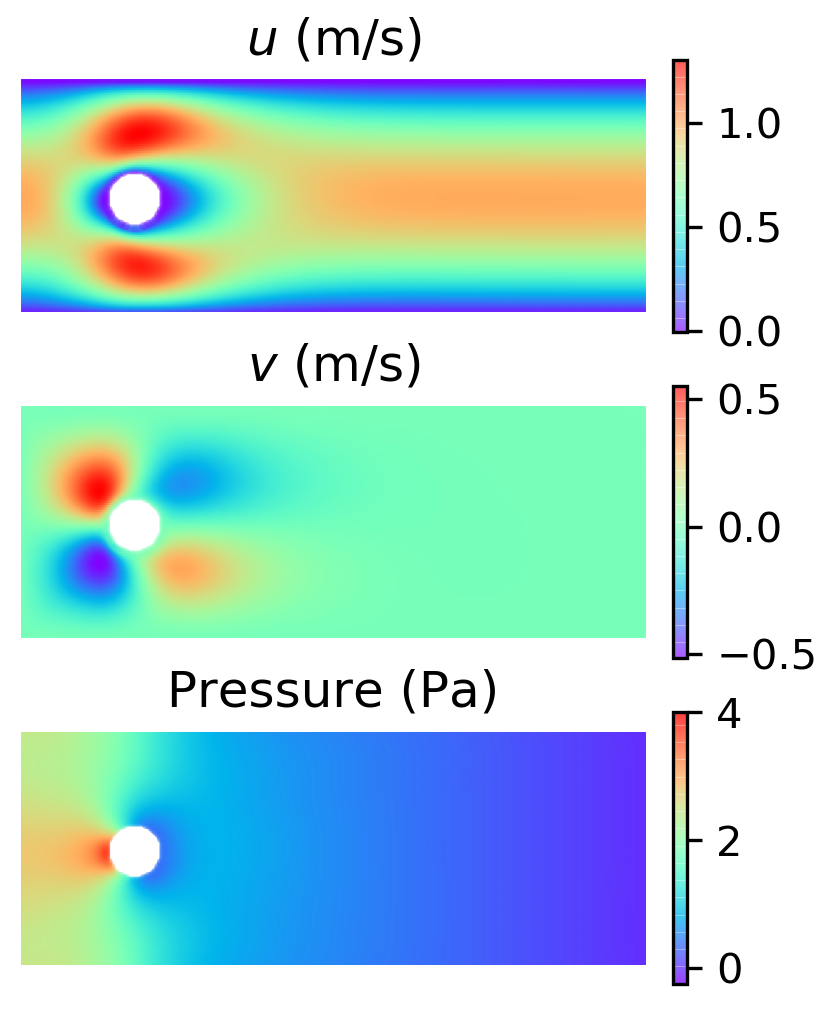}
}%
\subfigure[]{
\centering
\includegraphics[width=0.25\linewidth]{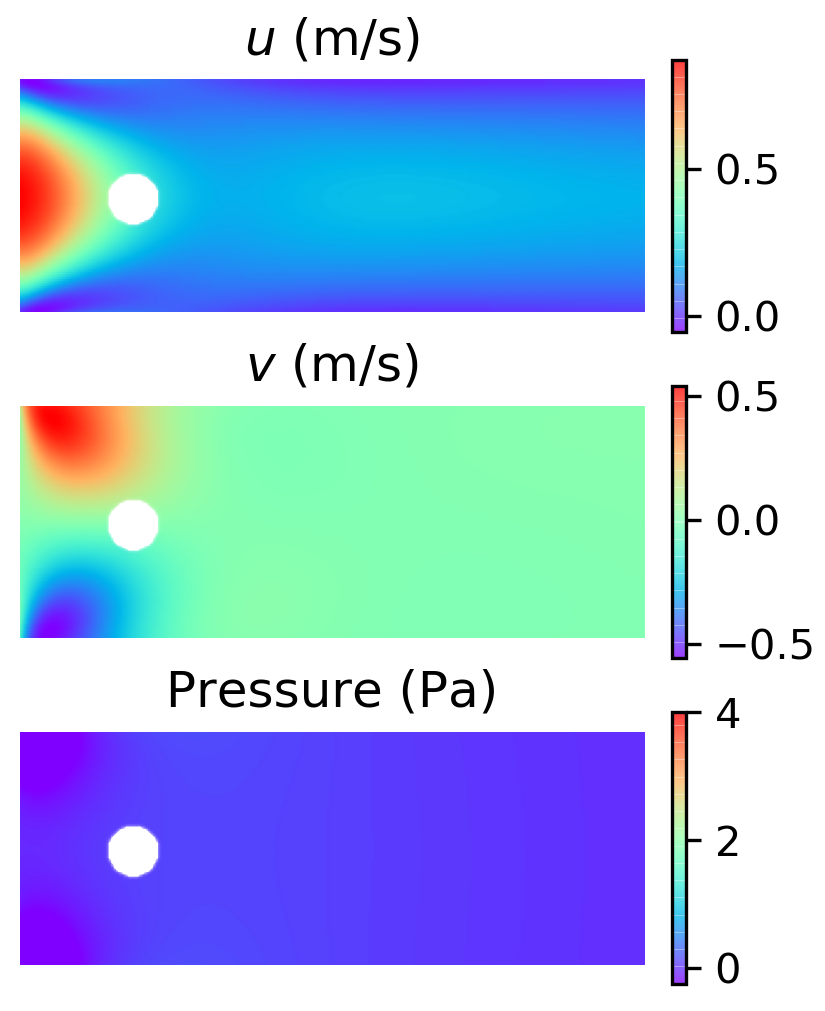}
}%
\label{steadycomp}
\caption{\color{black} Velocity and pressure fields of the steady flow passing a circular cylinder: (a) Reference solution from ANSYS Fluent; (b) Mixed-variable scheme solution with 8$\times$40 network; (c) Traditional scheme solution with 8$\times$40 network. The hyperparameters and collocation points for training these two PINNs are kept same.}
\end{figure*}

\subsection{Steady flow}
For the steady case, the dynamic viscosity and density of the fluid is $2\times10^{-2} \mathrm{kg/(m\cdot s)}$ and $1 \mathrm{kg/m^3}$ respectively. The normal velocity profile is defined as
\begin{equation} 
\label{velprofile} 
\begin{aligned}
u(0,y)&=4\mathrm{U_{max}}(\mathrm{H}-y)y/\mathrm{H}^2\\
\end{aligned}
\end{equation}
with $\mathrm{U_{max}}$ equal to {\color{black}1.0 m/s} which results in a small Reynolds number so that the flow is dominated by laminar flow. {\color{black} A total number of $N_g=50000$ collocation points, which includes $N_{db}=1200$ Dirichlet boundary (cylinder, wall, inlet) points and $N_{nb}=200$ Neumann boundary (outlet) points, are generated using Latin hypercube sampling (LHS) for the training the network. It should be noted that the collocation points are refined near the cylinder to better capture the details of the flow. A grid search strategy is used to find an optimal combination of depth and width for the network. The relative $\ell_2$ error defined by 
\begin{equation} 
\label{relativeL2E} 
\begin{aligned}
\epsilon = \frac{\sqrt{\sum_{i=1}^{M}||\mathbf{f}_{\text{pred}}^i-\mathbf{f}_{\text{ref}}^i||^2}}{\sqrt{\sum_{i=1}^{M}||\mathbf{f}_{\text{ref}}^i||^2}}
\end{aligned}
\end{equation}
is used as the metric for comparison, where $\mathbf{f}$ is the physical quantity of interest, and $M$ is the total number of reference points.} The Adam \cite{kingma2014adam} and Limited-memory BFGS (L-BFGS) optimizer \cite{Byrd1995} is employed to train the DNN due to their good convergence speed demonstrated in the tests. {\color{black}We also implement the traditional scheme for fluid dynamics employed in \cite{raissi2019physics, raissi2019deep} where the stream function and pressure $\{\psi,~p\}$ act as the output variables.} {\color{black} From the relative $\ell_2$ errors of the velocity field (see Table \ref{convtest1}), it can be seen that the network of $8\times40$ achieves the best result among all the configurations. The mixed-variable PINN improves the accuracy of numerical results over the traditional PINN. }

\begin{table}[h]
\centering
\caption{{\color{black} Relative $\ell_2$ errors (unit: $10^{-2}$) of the velocity field for different DNN configurations with $\beta=2$ (left: the traditional scheme; right: the proposed mixed-variable scheme).}}
\vspace{6pt}
\small
{\color{black}
\begin{tabular}{|l|c|c|c|}\hline
\diagbox[width=7em]{Width}{Depth}& 6 & 7 & 8  \\ \hline
40& 76.1/4.2 & 69.1/4.7 & 73.8/1.8 \\ \hline
50& 74.0/2.6 & 81.1/2.4 & 76.9/2.3\\ \hline
60& 74.5/2.3 & 75.2/2.4 & 75.4/1.9 \\ \hline
\end{tabular}\label{convtest1}}
\normalsize
\end{table}


\begin{figure}[t!]
	\centering
	\includegraphics[width=0.42\textwidth]{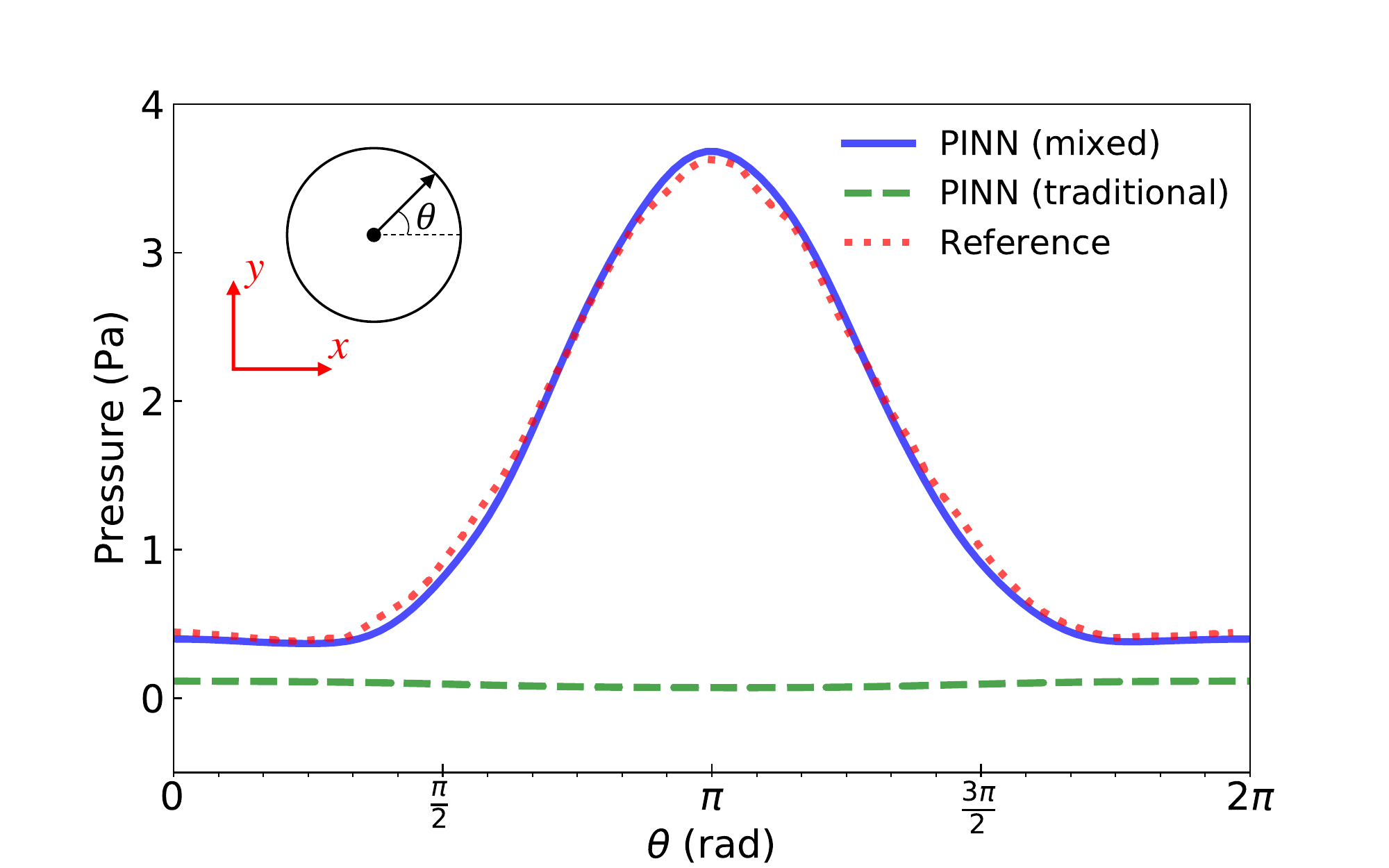}
	\caption{Distribution of pressure on cylinder. Network of 8$\times$40 and $\beta=2$ are used.}
	\label{presdistr}
\end{figure}

\begin{figure}[t!]
	\centering
	\includegraphics[width=0.45\textwidth]{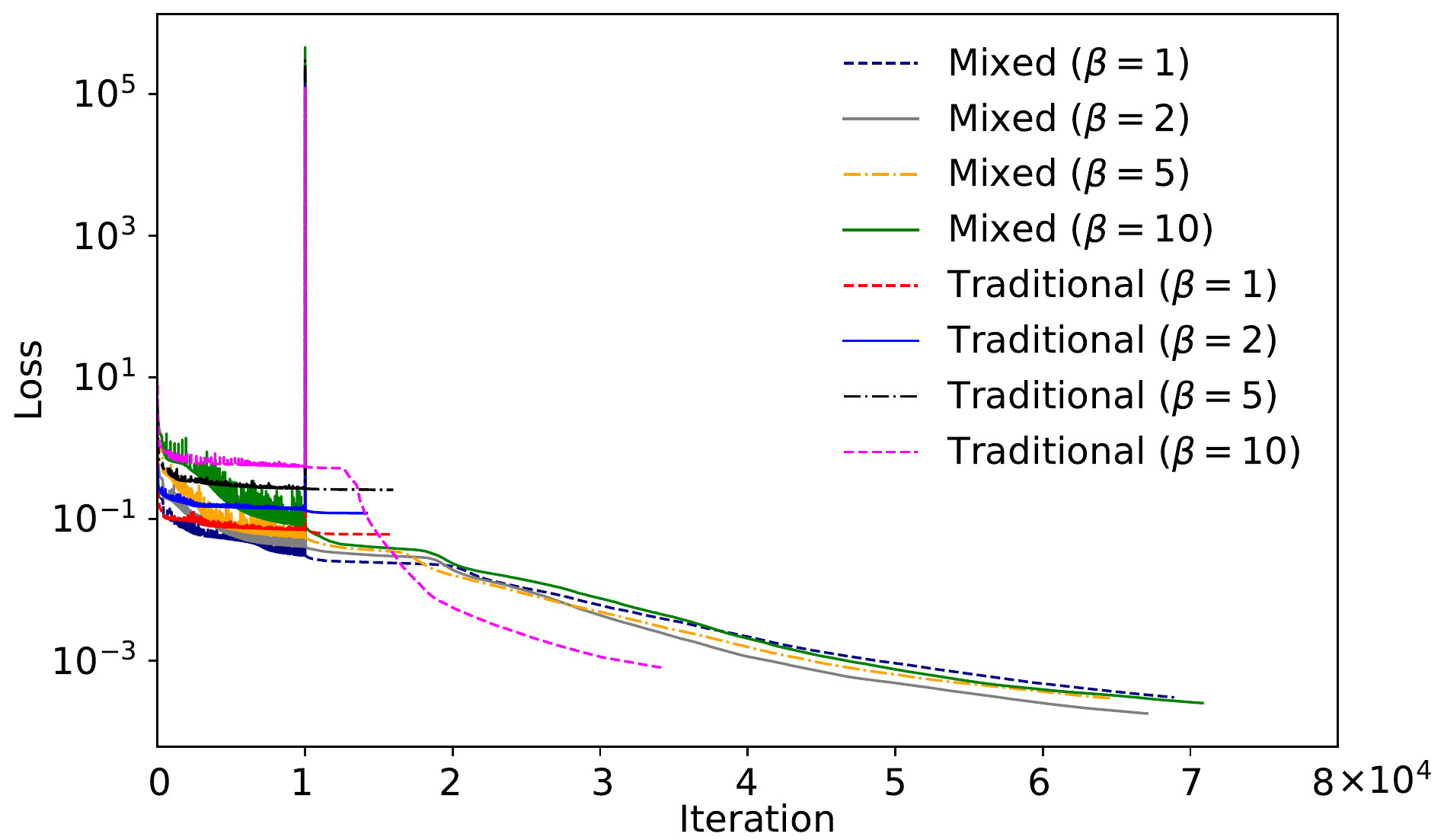}
	\caption{Comparison of convergence curves with respect to coefficient $\beta$. Network of $8\times40$ is used in all the cases. 10000 iterations trained with Adam optimizer followed by L-BFGS optimizer.}
	\label{eff_beta}
\end{figure}

\begin{figure}[t!]
	\centering
	\includegraphics[width=0.35\textwidth]{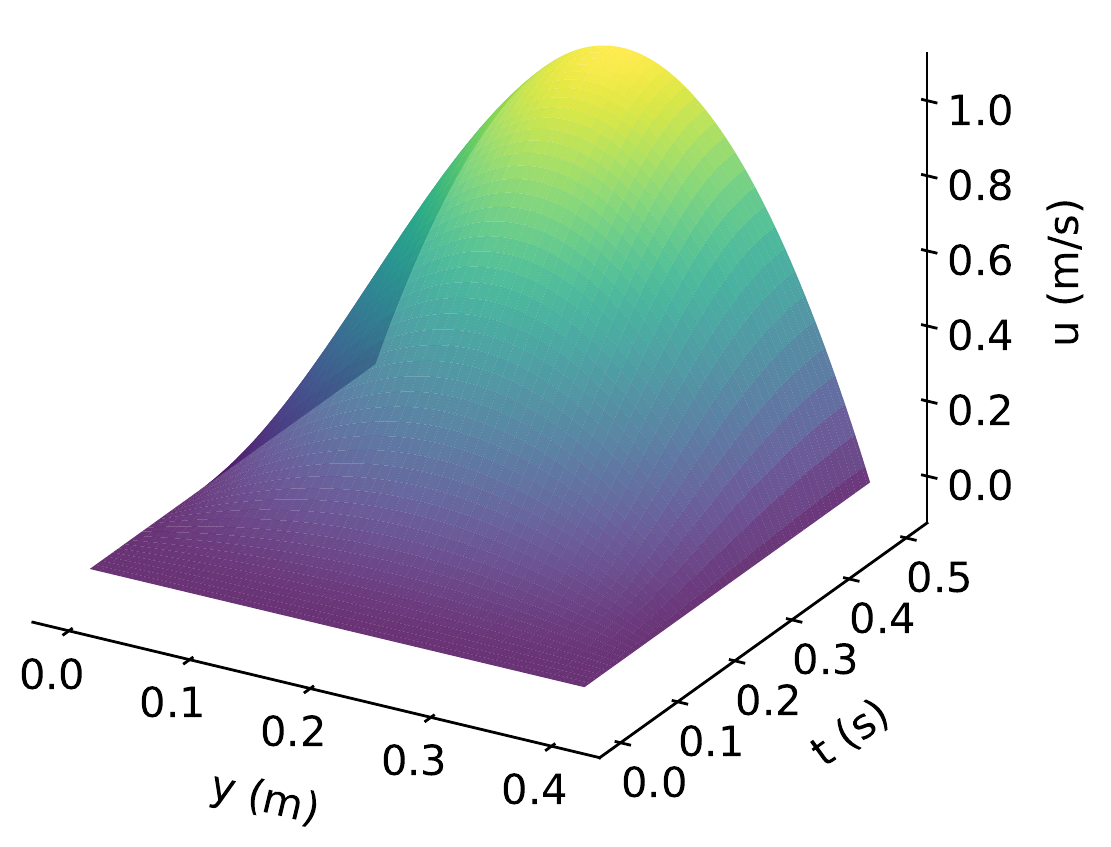}
	\caption{Transient normal velocity profile.}
	\label{visualvel}
\end{figure}

\begin{figure*}[t!]
    \centering
    \subfigure[t=0.3 s]{
    \centering
    \includegraphics[width=0.25\linewidth]{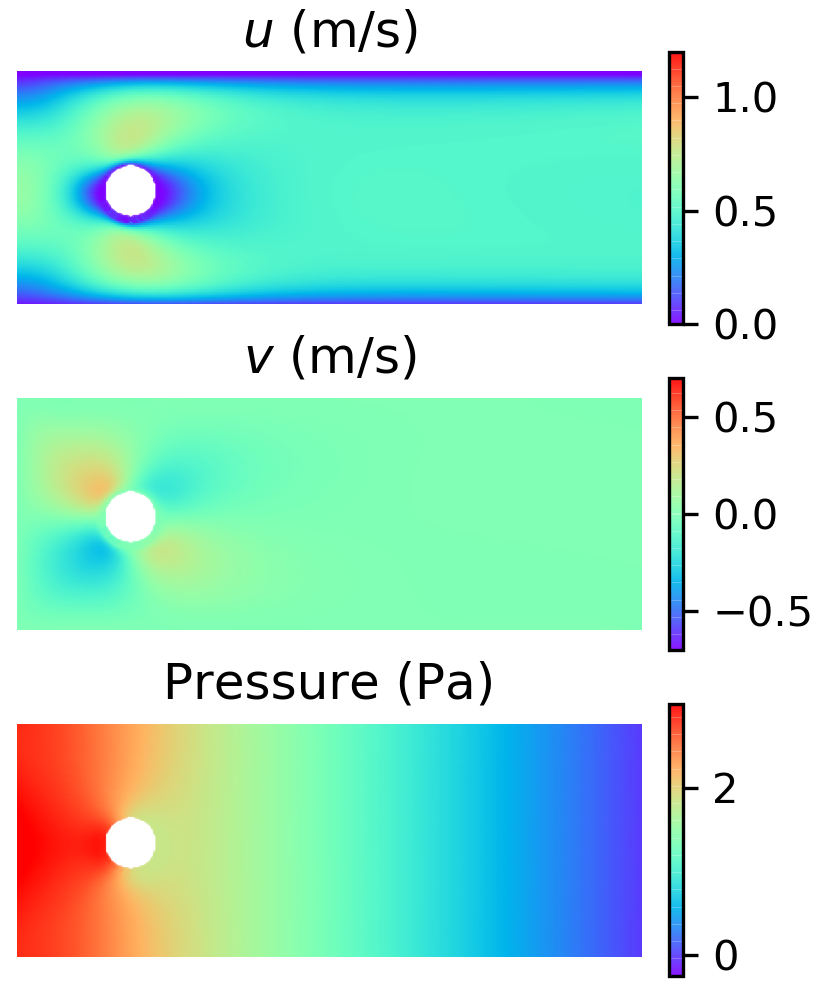}
    }%
    \subfigure[t=0.4 s]{
    \centering
    \includegraphics[width=0.25\linewidth]{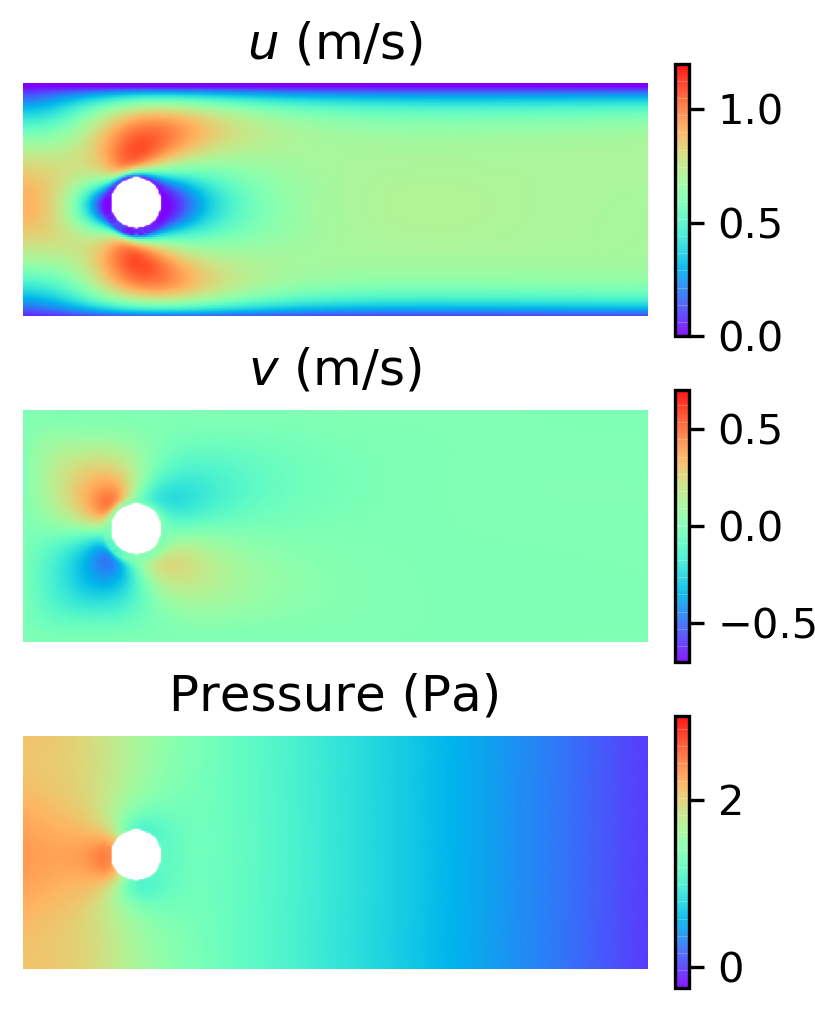}
    }%
    \subfigure[t=0.5 s]{
    \centering
    \includegraphics[width=0.25\linewidth]{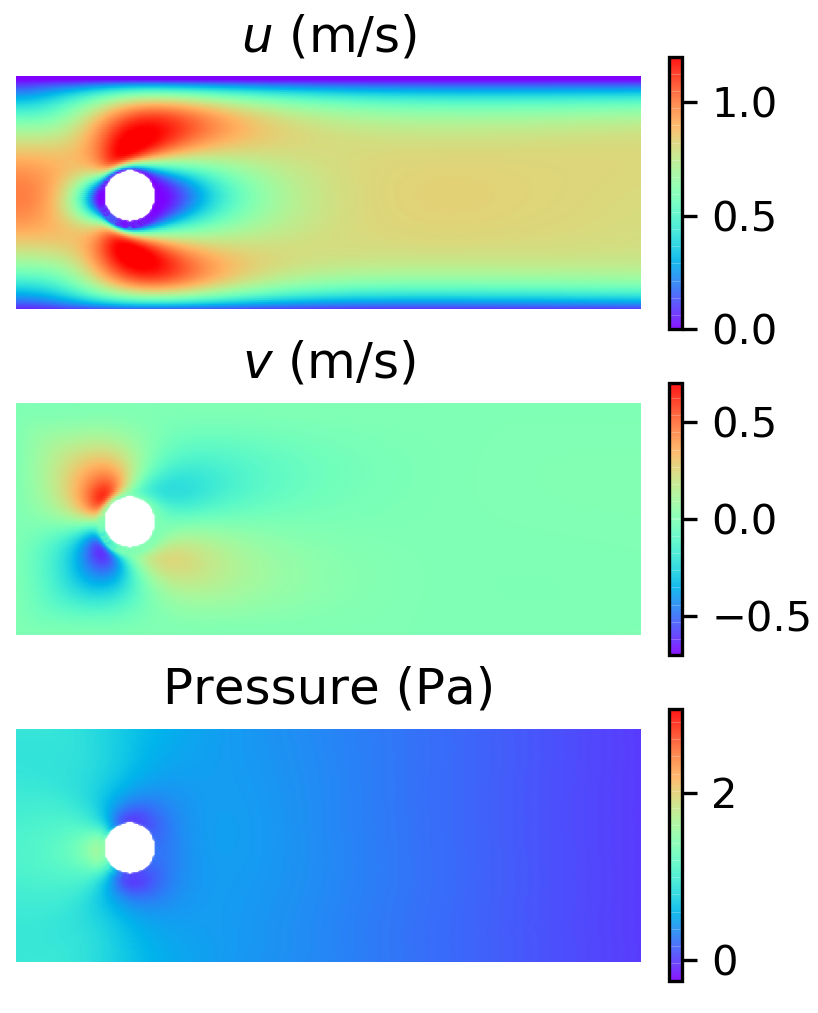}
    }
    \caption{Snapshots of the PINN-predicted transient flow fields passing a circular cylinder}
    \label{unsteadyuvp}
\end{figure*}

\begin{figure*}[h!]
	\centering
	\includegraphics[width=0.9\textwidth]{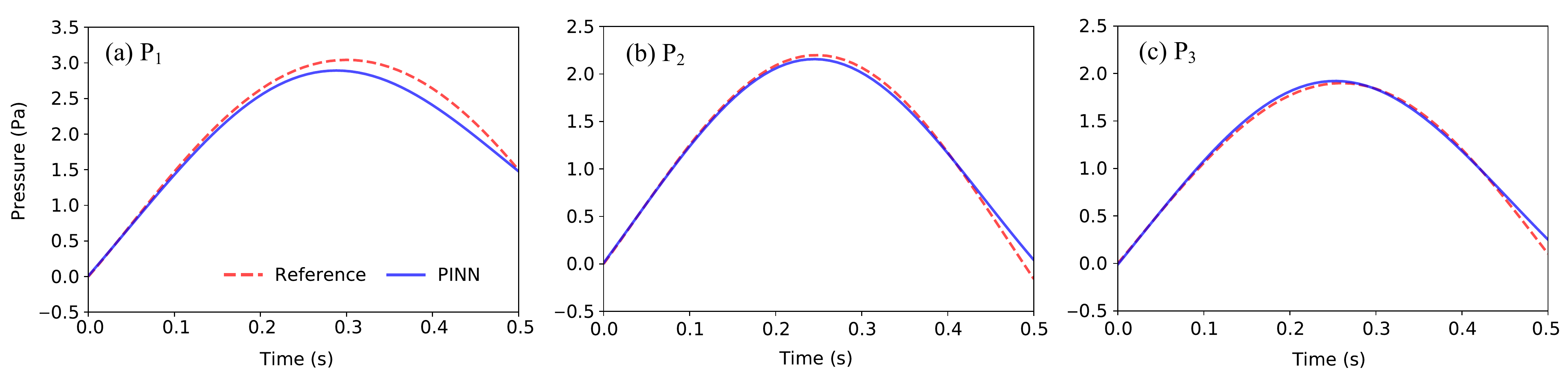}
	\caption{Pressure time histories on P$_1$, P$_2$ and P$_3$ probes.}
	\label{unsteadycomp}
\end{figure*}

The predicted velocity and pressure fields by the PINNs with mixed-variable and traditional scheme are shown in Fig. \ref{steadycomp}(b) and (c) respectively. The reference solution is obtained from the ANSYS Fluent 18.1 package (finite volume-based) \cite{fluent2011ansys} (see Fig. \ref{steadycomp}(a)). {\color{black}It can be observed that the PINN with the traditional scheme fails to model the flow.} {\color{black} In particular, the traditional scheme fails to enforce the non-slip condition on the lower and upper boundaries.} However, the steady velocity and pressure fields are well reproduced by the PINN with mixed-variable scheme. It is worth mentioning that the pressure distribution on the cylinder surface is typically of interest for computing the resultant drag and lift forces. Therefore, we compare the pressure distributions obtained by two types of PINN and ANSYS Fluent as shown in Fig. \ref{presdistr}. The overall agreement between the mixed-variable PINN and ANSYS Fluent is very good. 

{\color{black} We also compare the performance of these two schemes under various $\beta$ which controls the weight of the boundary condition loss. As shown in Fig. \ref{eff_beta}, the convergence of the traditional scheme is significantly affected by $\beta$, though the final loss can be reduced by increasing the value of $\beta$ up to 10. However, the mixed-variable scheme yields consistent results for various $\beta$. The improvement by the mixed-variable scheme is thanks to the reduced order of derivatives (see Section \ref{method}) required to construct the loss function, in comparison with the traditional scheme \cite{raissi2019physics, raissi2019deep}, which makes the optimization problem easier.}

\subsection{Transient flow}
The transient flow with the same computation domain depicted in Fig. \ref{diagram} is considered in this case. The dynamic viscosity of the flow is $\mu=5\times10^{-3} \mathrm{kg/(m\cdot s)}$ while the density is $\rho=1 \mathrm{kg/m^3}$. The time duration for the modeling is 0.5 s. Three virtual pressure probes P$_1$(0.15, 0.2) m, P$_2$(0.2, 0.25) m and P$_3$(0.25, 0.2) m are installed on the surface of the cylinder. The flow is initially still while a time-varying parabolic inlet velocity profile is applied subsequently, which is defined as
\begin{equation} 
\label{tranvelprofile} 
\begin{aligned}
u(0,y)&=4\left [\mathrm{sin}\left(\frac{2\pi t}{\mathrm{T}}+\frac{3\pi}{2}\right)+1\right ]\mathrm{U_{max}}(\mathrm{H}-y)y/\mathrm{H}^2\\
\end{aligned}
\end{equation}
where $\mathrm{U_{max}}$ equals to 0.5 m/s and the period $\mathrm{T}$ is 1.0 s. The remaining boundary conditions are the same as those in the previous example. The inflow velocity as a function of $t$ and $y$ is visualized in Fig. \ref{visualvel}. The width and depth of the network are selected to be 50 and 7 respectively while the coefficient $\beta$ is set to be 2. {\color{black}A total number of $N_g=120000$ collocation points, which include $N_{db}=9600$ points on cylinder, wall and inlet boundaries, $N_{nb}=3200$ points on outlet, and $N_{I}=3500$ points at initial time, are used to train the network.}
 
Three snapshots of the predicted flow fields are presented in Fig. \ref{unsteadyuvp} which shows the evolution of the flow as the inlet velocity increases over time. The reference flows obtained by ANSYS Fluent are {\color{black}not} shown here since the PINN-predicted result matches
very well with them. The pressure time histories on three probes obtained from the proposed PINN are compared with those from ANSYS Fluent, as depicted in Fig. \ref{unsteadycomp}. It can be seen that the proposed PINN approach can well predict the pressure time histories in a transient flow.

\section{Conclusions}\label{conc}
We propose a mixed-variable PINN scheme for modeling fluid flows, with particular applications to incompressible laminar flows. The salient features of the proposed scheme include (1) employing the general continuum equations together with the material constitutive law rather than the derived Navier-Stokes equations, and (2) using stream function to ensure the divergence free condition of the flow in a mixed-variable setting. {\color{black}The comparison study indicates the benefits (high accuracy and good trainability) of the proposed mixed-variable scheme.} In both the steady and transient flow cases, the result produced by the PINN shows a good agreement with the reference numerical solutions. 

It is notable that the applications in this paper are limited to the laminar flows at low Reynolds numbers, although the approach is in theory applicable to turbulent flows at large Reynolds numbers. However, it requires discretizing the computation domain with much finer collocation points which will lead to computer memory issues and drastically increase the computational cost. {\color{black}Our future work aims to address this challenge by developing a ``divide-and-conquer'' training scheme in the context of transfer learning, that is to divide the time domain into multiple steps and re-train the network partially while fixing the weights and the biases from the previous step \cite{goswami2020transfer}.}


\bibliographystyle{elsarticle-num}
\bibliography{refs}

\end{document}